\documentclass[
 prapplied,
 amsmath,amssymb,
 reprint, longbibliography, superscriptaddress
]{revtex4-2}

\usepackage{graphicx}
\usepackage{dcolumn}
\usepackage{bm}
\usepackage[utf8]{inputenc}
\usepackage[T1]{fontenc}
\usepackage{mathptmx}
\usepackage{etoolbox}
\usepackage{amssymb}
\usepackage{amsmath}
\usepackage{hyperref}
\usepackage[version=3]{mhchem}
\usepackage{esint}
\usepackage{gensymb}
\usepackage{url}
\usepackage{mathptmx}
\usepackage{siunitx}

\begin{document}

\preprint{APS/123-QED}


\title[Surface and Bulk Two-Level System Losses in Lithium Niobate Acoustic Resonators]{Surface and Bulk Two-Level System Losses in Lithium Niobate Acoustic Resonators}

\author{Rachel G. Gruenke-Freudenstein}
\email{rgruenke@stanford.edu}
\author{Erik Szakiel}
\author{Gitanjali P. Multani}
\author{Takuma Makihara}
\author{Akasha G. Hayden}
\author{Ali Khalatpour}
\affiliation{ 
Department of Applied Physics and Ginzton Laboratory, Stanford University, 348 Via Pueblo Mall, Stanford, California 94305, USA
}
\author{E. Alex Wollack}
\affiliation{AWS Center for Quantum Computing, Pasadena, California 91106, USA}

\author{Antonia Akoto-Yeboah}\affiliation{Department of Materials Science and Engineering,
The Ohio State University, Columbus, Ohio 43210, USA}

\author{Salva Salmani-Rezaie}
\affiliation{Department of Materials Science and Engineering,
The Ohio State University, Columbus, Ohio 43210, USA}
\author{Amir H. Safavi-Naeini}
\email{safavi@stanford.edu}
\affiliation{ 
Department of Applied Physics and Ginzton Laboratory, Stanford University, 348 Via Pueblo Mall, Stanford, California 94305, USA
}

\date{\today}

\begin{abstract}
Lithium niobate (LN) is a promising material for building acoustic resonators used in quantum applications, but its performance is limited by poorly understood material defects called two-level systems (TLS). In this work, we fabricate high-performance acoustic resonators from LN with quality factors up to $6\times10^7$ and use them to separate bulk and surface contributions to TLS loss. By comparing these bulk acoustic wave (BAW) resonators with previous surface acoustic wave and phononic crystal studies, we show that devices with high surface participation ratios are limited by surface TLS, while our BAW devices reveal an intrinsic bulk TLS limit. Through systematic surface treatments and microscopy, we demonstrate that BAW resonator performance remains unchanged despite surface modifications, confirming operation in a bulk-limited regime. Our work establishes quantitative bounds on both surface and bulk TLS losses in LN, within the context of material growth and fabrication approaches we have pursued, and provides guidance for future device engineering and materials development.
\end{abstract}

\maketitle

\section{Introduction}

Acoustic resonators such as phononic crystals (PNCs) \cite{arrangoiz2019resolving, wollack2022quantum,lee2023strong,emser2024thin}, surface acoustic wave resonators (SAWs) \cite{satzinger2018quantum,sletten2019resolving,bienfait2019phonon}, and bulk acoustic wave resonators (BAWs) \cite{chu2018creation, von2022parity, bild2023schrodinger, campbell2024low, diamandi2024quantum, franse2024high} have emerged as platforms for quantum science and technology over the last decade. As quantum acoustics experiments mature with more stringent requirements, higher fidelity quantum protocols using acoustic resonators will be required. Higher fidelity invariably requires large interaction rates and low dissipation. Lithium niobate (LN) is one of the most piezoelectric low loss materials and, therefore, enables strong interactions with microwave fields. By understanding the loss channels and mitigating them, we enable development of high cooperativity devices. At cryogenic temperatures, lithium niobate acoustic resonators such as PNCs or SAWs suffer from significant two-level system (TLS) losses \cite{wollack2021loss, gruenke2024surface}. While there has been substantial progress in understanding TLS across many platforms \cite{zhang2024acceptor, martinis2005decoherence, muller2019towards, crowley2023disentangling}, key questions remain about their nature in LN. In this work, we fabricate ultra-low loss bulk acoustic wave (BAW) resonators on X-cut LN, with (high-power) internal quality factors up to $60\times10^6$. While these devices point to a promising path for highly coherent quantum acoustics, they are also a powerful tool for studying LN TLS. We show that these devices are still limited by TLS losses, but that unlike previously studied lithium niobate PNCs and SAWs, the losses are insensitive to surface treatments. By comparing the surface and bulk strain distribution of PNCs, SAWs and BAWs and their measured TLS losses, we extract the contributions of surface- and bulk-lying TLS in LN. These results provide a picture on how to engineer acoustic strain to reduce TLS loss. Using measurements of TLS participation in different device regions, combined with surface analysis from AFM and TEM, we identify likely sources of TLS in LN and propose future studies to better understand these loss mechanisms.

\section{Surface and Bulk TLS Losses in Lithium Niobate Acoustic Resonators}

By studying TLS-induced losses and frequency shifts  in different acoustic resonators  that have undergone similar fabrication processes, our goal is to identify how TLS couple to strain as a function of the resonator geometry and therefore infer where TLS are most likely present. Our approach is to first extract TLS properties using temperature-dependent resonant frequency shifts or power saturated losses. These measurements provide us with the $F \delta^0_{\text{TLS}}$ parameter, where $\delta^0_{\text{TLS}}$ is the loss tangent associated with the TLS and $F$ is the filling fraction representing the fraction of the mode's energy that interacts with the TLS. Though a single measurement on a sample with many TLS does not have enough information to let us distinguish between $F$ and $\delta^0_{\text{TLS}}$, by combining measurements across many devices and mode profiles that have different $F$ we can begin to disentangle these two parameters.

\begin{figure}[h]
\includegraphics[width=\linewidth]{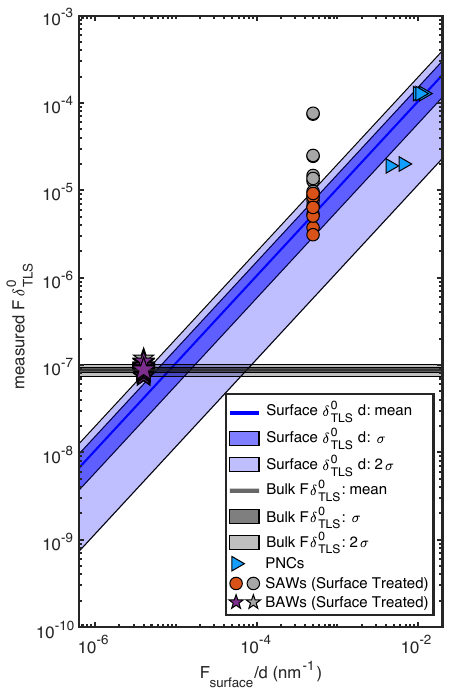}
\caption{\label{fig:fig_Fdel0} \emph{Lithium Niobate TLS Losses} We show a summary of TLS loss measurements on three LN acoustic resonators with varying length scales and strain profiles. Each resonator is characterized by its measured $F\delta^0_{\text{TLS}}$ and the fractional acoustic mode energy in a small surface layer of depth $d$ normalized by the depth, or $F_\text{surface} /d$. By comparing these results, we show how the TLS losses scale with surface participation of the acoustic strain. The measured $F\delta^0_{\text{TLS}}$ drops off with smaller surface participation, suggesting dominating surface-lying TLS. Blue lines represent contours of constant surface $\delta^0_{\text{TLS}} \times d$ ($1.05\times10^{-2} \pm 4.7\times10^{-3}$) for standard LN processed devices, determined from the PNC and SAW values. We see that the BAW data falls outside of this range and has little variation with surface treatment. From the BAW data, we determine CLN bulk TLS losses represented by gray contours of constant $F\delta^0_{\text{TLS}}$ ($8.78\times10^{-8} \pm 6.7\times10^{-9}$).}
\end{figure}

To understand the nature of TLS losses in our acoustic devices, we analyze the $F \delta^0_{\text{TLS}}$ measurements across different resonator geometries (Figure~\ref{fig:fig_Fdel0}). We combine data spanning nearly four orders of magnitude in this plot, from our new BAW resonator measurements to studies of PNCs \cite{wollack2021loss} and SAWs \cite{gruenke2024surface}. A clear trend emerges: the measured $F \delta^0_{\text{TLS}}$ decreases as we move from PNCs to SAWs to BAWs, suggesting reduced surface participation leads to reduced TLS coupling.
To quantify this relationship, we define a surface filling fraction $F_\text{surface}$ that captures how much of the acoustic energy interacts with surface-lying TLS. The filling fraction is generally given by:
\begin{align}\label{eq:fill_gen} F = \frac{\int_{V_h} S_I c_{IJ} S_J dV}{\int_{V}   S_I c_{IJ} S_J dV}, \end{align}
where $c$ is the stiffness tensor, $S$ is the acoustic strain, $V$ is the total resonator volume, and $V_h$ is the volume containing TLS \cite{gao2008physics, auld1973acoustic}. For surface-lying TLS, we assume a thin layer of depth $d$ at all exposed surfaces hosts the TLS. For very thin surface layers (where thickness $d$ is much smaller than the acoustic wavelength), we find that $F$ scales linearly with $d$, so $F/d$ becomes a constant. This lets us characterize the surface effects independently of the exact surface layer thickness. We calculate this quantity analytically for our BAW resonators by assuming a periodic strain along the cavity length (see Appendix~\ref{calc_F}). For the PNCs and SAWs, we calculate this quantity by solving for the mode strain distribution and performing the integration in COMSOL \cite{COMSOL}. The PNCs have the largest surface filling fraction as they are fabricated on thin film LN of around 250nm thickness. The BAWs have the smallest surface filling fraction as their strain profile extends uniformly through the bulk of the crystal, 500 $\mu$m thick. SAWs have a surface filling fraction between the PNCs and BAWs. 

By plotting $F_\text{surface}/d$ versus measured $F\delta^0_{\text{TLS}}$ for all resonators, we observe that TLS losses generally decrease when we reduce surface participation. By examining our previously studied PNCs and SAWs that underwent standard LN processing, we can establish bounds for surface-related TLS losses, finding that the product of loss tangent times damage layer depth falls between $1.05\times10^{-2} \pm 4.7\times10^{-3}$. Notably, the BAW measurements sit above this trend line, showing higher TLS losses than would be predicted from surface effects alone and demonstrating insensitivity to surface treatments (see Section~\ref{BAW_TLS}). This suggests bulk rather than surface TLS losses dominate their $F\delta^0_{\text{TLS}}$ values. We hypothesize these bulk losses may be related to intrinsic defects in congruently Czochrlaski-grown LN, which exhibits lithium deficiency and associated niobium antisite defects. Based on our measurements, we estimate the bulk TLS losses correspond to an $F\delta^0_{\text{TLS}}$ of $8.78\times10^{-8} \pm 6.7\times10^{-9}$. 

These findings provide a clear strategy for minimizing TLS losses in future LN devices. By measuring $F\delta^0_{\text{TLS}}$ and calculating $F_\text{surface} /d$, we can use the information in Figure~\ref{fig:fig_Fdel0} to determine whether a device is limited by surface or bulk TLS effects. For surface-limited devices, surface treatments may be able to reduce the $\delta^0_{\text{TLS}} d$ product, while for bulk-limited devices, addressing material defects through techniques like vapor transport equilibration may lead to higher coherence devices. In the future, studies exploring bulk LN acoustic resonators with $F_\text{surface}/d$ values outside our current range (below $10^{-6}$ or above $5\times10^{-5}$) will further solidify our understanding of these loss mechanisms.

\section{BAW Design and Total Losses}

\begin{figure}[h!]
\includegraphics[width=\linewidth]{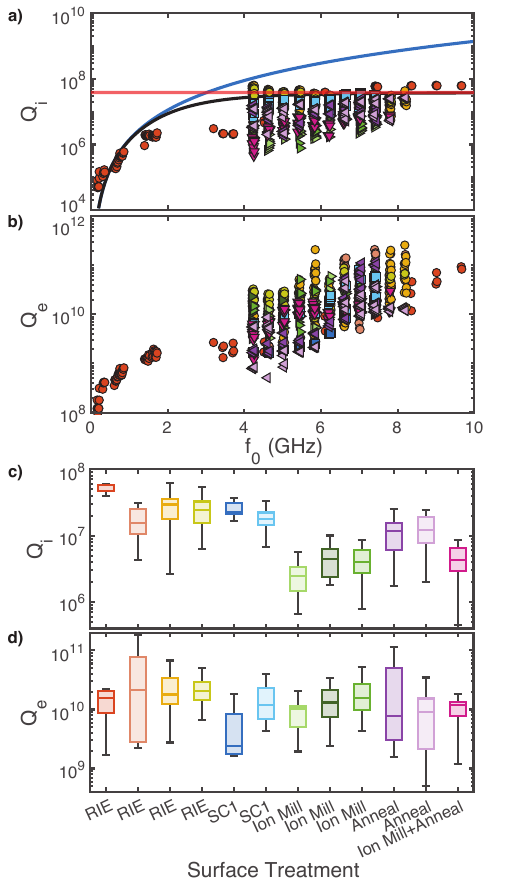}
\caption{\label{fig:fig_losses} \emph{Overview of the BAW  total losses} We show the measured high power ($10^8 < \langle n \rangle < 10^9$) internal (a) and external (b) quality factors of each surface treated BAW device. While most devices are measured in a 4-8GHz stopband set by our circulators, one device is measured without circulators, allowing us to measure across the full stopband. $Q_i$ and $Q_e$ increase with frequency, though $Q_i$ hits a maximum value of $6\times10^7$ and stops increasing after 8 GHz. We model the total losses (black) as a sum of diffraction losses (blue) dominating at low frequencies and TLS losses (red) dominating at high frequencies. Box plots summarizing all device data measured between 4-8~GHz is shown in (c) and (d). While $Q_e$ is surface treatment insensitive, $Q_i$ is reduced in our "Ion mill" and "Ion mill+Anneal" devices.}
\end{figure}

\begin{figure*}[ht]
\includegraphics[width=\linewidth]{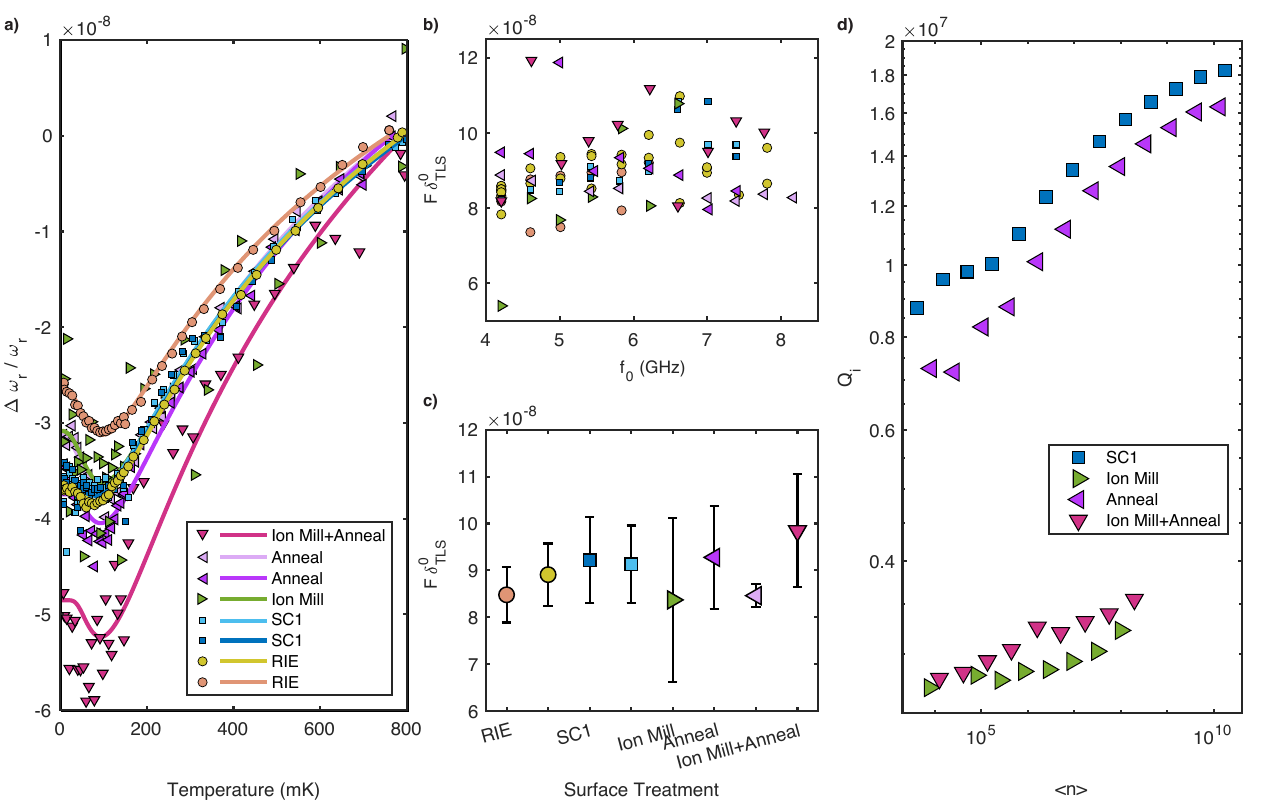}
\caption{\label{fig:fig_tls} \emph{Overview of the BAW TLS losses} (a) Temperature swept resonance frequency shift of all surface treated devices for a starting resonance near 4.6~GHz. The reference frequency is set as the frequency at 800~mK. (b) Extracted $F\delta^0_{\text{TLS}}$ from temperature swept measurements of resonances from 4-8GHz. (c) A summary of all fitted $F\delta^0_{\text{TLS}}$ for each surface treated measured device. Each resonator is represented by the mean of all fitted resonances of the unique device in the 4-8GHz band. Error bars are set by the standard deviation. (d) Measured $Q_i$ for decreasing internal resonator power. We see that while all resonators have decreasing $Q_i$ with power, the "Ion mill" and "Ion mill + Anneal" devices have a much lower overall $Q_i$, demonstrating ion mill induced non-TLS losses.}
\end{figure*}

\begin{figure*}[ht]
\includegraphics[width=\linewidth]{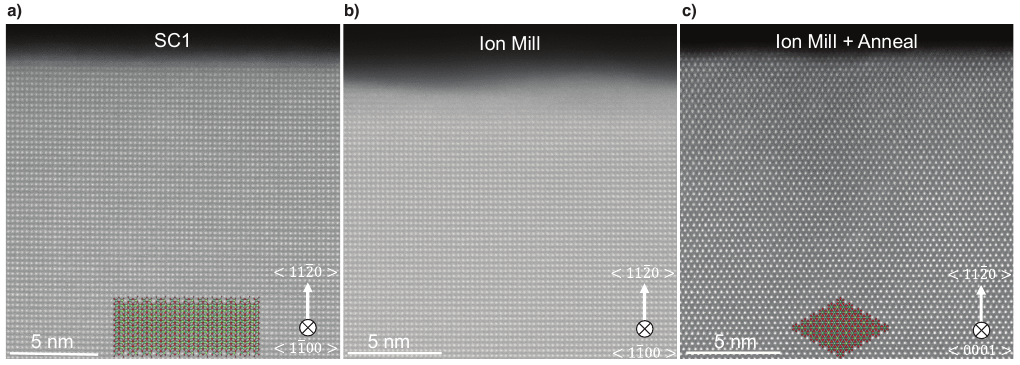}
\caption{\label{fig:fig_tem} \emph{LN TEMs} HAADF-STEM images of cross-sectioned samples for (a) "SC1", (b) "Ion Mill" and (c) "Ion Mill+Anneal" lithium niobate along the $\langle 1 \overline{1} 0 0 \rangle$ and $\langle 0001 \rangle$ zone axes, respectively. The "Ion mill" sample shows $\approx3$ nm thick amorphous layer at the surface, indicative of ion-induced damage. After annealing, the damaged layer is removed, and the surface regains a more ordered structure.}
\end{figure*}

\begin{figure}[h!]
\includegraphics[width=\linewidth]{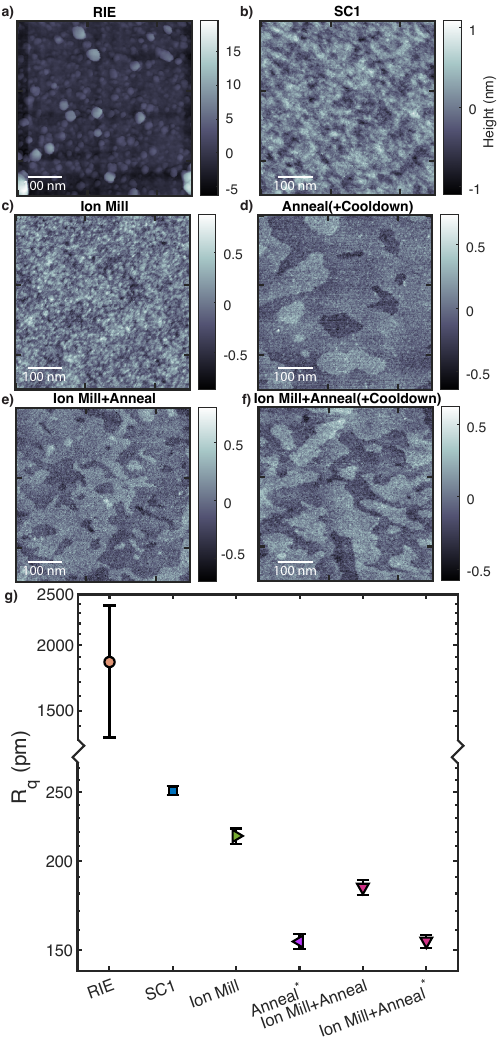}
\caption{\label{fig:fig_afm} \emph{LN AFMs} Images (a) - (f) show surface height maps taken with PeakForce AFM. Morphology patterns of "SC1" and "Ion mill" devices resemble chemical-mechanical polished LN, while the "RIE" device shows significant surface debris, and "Anneal" and "Ion Mill+Anneal" devices show atomically flat terraces. Roughness of each device (averaged from several scans) are shown in (g). Error bars shown are calculated from the standard deviation of all scans of the same device. A dramatic drop in surface roughness occurs after SC1 cleaning the RIE stripped substrate. Annealing also slightly reduces the roughness. Asterisk on surface treatment labels correspond to a cooldown after surface treatment but before AFM measurement.}
\end{figure}

We now turn to our experimental investigation using bulk acoustic wave resonators, the data obtained from which we have already shown in Figure~\ref{fig:fig_Fdel0}. These devices offer an important platform for studying TLS effects due to their minimal surface participation and high quality factors.

We fabricate ultra-low loss bulk acoustic wave resonators using flip-chip integration. The BAW cavity is a flat-flat mirror cavity. We strip the 500~{\textmu}m X-cut LN of a 1~{\textmu}m layer of protective silicon using a 1:7 $\text{SF}_\text{6}$:Argon RIE etch \cite{minnick2013optimum}. While we also developed a process to make a parabolic hard mask for etching lensed mirrors \cite{kharel2018ultra, khonina2024grayscale}, we found that we obtain the lowest loss devices with flat mirrors. The BAW cavity chip can then be surface treated with one of the four additional surface treatment studies we test, including argon ion milling, annealing in atmosphere at 500$^{\circ}$C for 72 hr, SC1 (10:1:1 water:ammonium hydroxide:hydrogen peroxide) base corrosive dip and piranha (3:1 sulfuric acid:hydrogen peroxide) acid corrosive dip. A list of all surface treatments is found in Table \ref{surface_treatment}. 

The BAWs are driven by capacitive non-galvanic coupling to a circular aluminum antenna. The substrate of the electrode chip is 425~{\textmu}m thick X-cut LN with a 1~{\textmu}m layer of protective silicon. The LN is thinned via backgrinding down 75~{\textmu}m to roughen the backside surface and change the acoustic free spectral range (FSR) of the modes it supports to avoid clashing with the BAW chip. There is also a layer of silicon on the electrode chip separating the electrodes from the LN to help reduce acoustic coupling into the electrode chip substrate. The antenna and feedlines are fabricated by liftoff of 75nm of aluminum. The two chips are then aligned with a Finetech Lambda flip-chip bonder, and secured with GE Varnish. The gap between the electrode and BAW cavity chips are maintained by 1~{\textmu}m aluminum spacers on the electrode chip, also fabricated by liftoff on the electrode chip. 

The advantage of flat-flat mirror cavities stems from their simplicity in design and processing which helps minimize surface damage leading to very low loss resonances \cite{franse2024high}. However the flat-flat cavities have no mechanism for transversal confinement of the mode and, therefore, have a dense mode spectrum with a forest of weakly coupled higher order spurious modes present. Nonetheless, the linewidth of each mode is sufficiently small, so they are individually distinguishable. Moreover, in contrast to SAW excitation which is usually narrowband due the periodic nature of the electrodes and mirrors, the BAW transducers excite modes over a very broad range of frequencies. Our devices cover the full measurement range spanning from a few MHz to nearly 10 GHz, with a FSR of 13.2 MHz. This FSR defines the distance between groupings of modes which likely have different transverse structure, but similar frequencies due to the flat-flat geometry. Within each grouping we see between 3 and 10 high quality factor resonances. The FSR is described well by the odd modes of longitudinal bulk speed of sound in 500~{\textmu}m thick X-cut LN -- ${v_l}/{L}$. See Appendix~\ref{flatflat} for details on calculating the longitudinal bulk speed and FSR from the Christoffel equation for X-cut LN. The large number of high-Q modes present makes our system ideal for gathering detailed statistics on loss mechanisms in LN.

The broadband nature of our measurements provides insight into many different loss mechanisms. By measuring a control device at high power and at cryogenic temperatures (10 mK) across frequencies from 6.6 MHz to 10 GHz, we observe that $Q_i$ decreases at lower frequencies, consistent with acoustic beam diffraction losses. Notably, we find no evidence of Rayleigh scattering, which would manifest as increased losses at higher frequencies. A detailed analysis of these diffraction losses is presented in Appendix~\ref{diffraction}.

\begin{table*}[t]
\caption{\emph{Surface Treatment Steps for Each Device} A summary of all surface treatments performed on the BAW substrates prior to measurement.}
\centering
\begin{ruledtabular}
\label{surface_treatment}
\begin{tabular}{l l}
\textbf{Device} & \textbf{Treatment Steps} \\
"RIE" & RIE etching + solvent cleaning \\
"SC1" & RIE etching + SC1 corrosive \\
"Ion Mill" & RIE etching + SC1 corrosive + Ar Ion Mill + SC1 corrosive \\
"Anneal" & RIE etching + SC1 corrosive + Anneal 500$^{\circ}$C 72hr + Piranha \\
"Ion Mill + Anneal" & RIE etching + SC1 corrosive + Ar Ion Mill + SC1 corrosive + Anneal 500$^{\circ}$C 72hr + Piranha \\
\end{tabular}
\end{ruledtabular}
\end{table*}

We compare the total internal quality factor for each surface treated device (see Figure \ref{fig:fig_losses}); we measure resonances at 10 mK between 4-8 GHz in each device set by two cryogenic circulators in the measurement chain. For our control devices (those with only RIE processing of Si and solvent cleaning), we measure median high-power quality factors between $15\times10^6$ and $57\times10^6$ in the 4-8 GHz band. Surface treatments have varying effects on these quality factors: devices that underwent ion milling show significantly reduced $Q_i$, with values between $2.4\times10^6$ and $4.4\times10^6$. This reduction persists even after annealing, with the "Ion mill+
Anneal" device showing $Q_i = 4.6\times10^6$. In contrast, devices treated with only SC1 cleaning or annealing maintain quality factors closer to the control devices, though annealed devices show a slight reduction in $Q_i$. These surface treatments do not alter the external coupling $Q_e$, and all devices remain undercoupled.

While these variations in quality factors across different surface treatments provide insight into process-dependent losses, they cannot directly distinguish between different loss mechanisms such as surface roughness or TLS. In the following section we pursue an approach to isolate and quantify the TLS effects independently of other loss mechanisms that may be introduced by our surface treatments.

\section{BAW TLS Losses} \label{BAW_TLS}
To characterize TLS losses in the resonators, we measure temperature-dependent resonance frequency shifts and power saturation of $Q_i$. For the temperature sweeps, we measure the resonance frequency of our acoustic modes from 10~mK to 800~mK, and extract the TLS loss $F \delta^0_{\text{TLS}}$ with 
\begin{align}\label{eq:one}
    \!\!\!\!\!\frac{\Delta\omega_r}{\omega_r} = \frac{F\delta_{\text{\text{TLS}}}^0}{\pi}\! \left[ \text{Re}\!\left\{\!\Psi\!\left(\frac{1}{2}\!+\!\frac{\hbar\omega_r}{2\pi i k_B T}\right)\!\right\} \!-\!\ln\frac{\hbar\omega_{r}}{2\pi k_B T}\right].\!\!\!\!\!\!
\end{align} 
Here, $\Delta \omega_r$ is the frequency shift of the mechanical oscillator from its nominal frequency $ \omega_r$ at 800~mK, and $T$ is the temperature \cite{gao2008physics}. We perform these measurements on approximately 10 resonances spaced by roughly 600~MHz between 4 and 8 GHz for each of the surface treated resonators as shown in Figure~\ref{fig:fig_tls} (a). We thus extract the frequency and surface treatment dependence of $F \delta^0_{\text{TLS}}$. We see very little frequency dependence from 4 to 8 GHz [Figure~\ref{fig:fig_tls} (b)]. We also see that the surface treatments cause very little change in the TLS loss averaged over the 4 GHz bandwidth [Figure~\ref{fig:fig_tls} (c)]. The mean $F\delta^0_{\text{TLS}}$ for all devices stays around $8.9\times10^{-8}$, and the largest measured $F\delta^0_{\text{TLS}}$ of $9.8\times10^{-8} \pm 1.2\times10^{-8}$(from the "Ion mill + Anneal" device) remains within the error range of the control sample ($8.7\times10^{-8} \pm 6\times10^{-9})$. 

Despite the significant differences in total quality factors, all devices exhibit similar TLS coupling strengths. To understand this apparent contradiction, we analyze power-dependent loss measurements shown in Figure~\ref{fig:fig_tls} (d). To separate TLS losses ($Q_{\text{TLS}}$) from other residual losses ($Q_{\text{i,res}}$), we use the following equation for TLS loss power saturation:  
\begin{equation}
\frac{1}{Q_{i,\text{tot}}} = \frac{F \delta^0_\text{\text{TLS}} \tanh(\frac{\hbar \omega_r}{2 k_B T})}{\sqrt{1 + \frac{\langle n \rangle}{n_c}^\beta}} + \frac{1}{Q_{i,\text{res}}}.
\label{eq:q_tls}
\end{equation}
Assuming $\beta=0.5$ and $n_c=10^6$, we find that the $Q_{i,\text{res}}$'s for "Ion mill" and "Ion mill + Anneal" devices are $3.5\times10^6$ and $3.9\times10^6$, respectively. In comparison, the "SC1" and "Anneal" devices which were not ion milled have $Q_{i,\text{res}} = 20.4\times10^6$ and  $18.8\times10^6$. However, all devices have a similar $Q_{\text{TLS}}$, ranging between $8.8\times10^6$ to $15\times10^6$. This analysis reveals that the quality factor degradation in ion-milled devices stems from non-TLS loss mechanisms, potentially due to increased conductivity in the LN caused by vacuum heating during the milling process.

\section{Lithium Niobate Metrology}

\subsection{Transmission Electron Microscopy}
To evaluate the atomic structure of the top $\approx 20$ nm of the LN, we image witness devices with similar surface treatment steps using scanning transmission electron microscopy (S/TEM). Cross-sectional specimens for S/TEM were prepared using a focused ion beam (FIB) system (FEI Helios NanoLab 600 DualBeam). Prior to FIB milling, the sample surface was coated with a thin carbon layer applied using a permanent marker (Sharpie) to protect it from beam-induced damage. The lamellae were then thinned and polished with Ga ions at 5 kV to minimize surface damage.
S/TEM imaging was performed on a Thermo Fisher Scientific Themis-Z microscope operated at an accelerating voltage of 200 keV and a semi-convergence angle of 30 mrad. High-angle annular dark-field (HAADF) images were acquired using a 64–200 mrad detector range. To achieve high-resolution images with improved signal-to-noise ratio, multiple rapid scans (2048 × 2048 pixels, 200 ns per pixel) were averaged.
The HAADF-STEM images of cross-sectioned lithium niobate samples reveal clear differences in surface structure following ion milling and subsequent annealing (see Figure \ref{fig:fig_tem}). The "Ion mill" sample has $\approx3$ nm thick amorphous layer at the surface, indicating significant damage caused by ion milling. After annealing, this amorphous layer is no longer visible, and the surface recovers a well-ordered, crystalline structure. A low-magnification STEM image covering a larger surface area, presented in Appendix Figure \ref{fig:si_tem_highres}, further confirms this recovery. 
When compared to the "SC1"-cleaned sample, the "Ion mill + Anneal" surface shows improved structural quality, with a higher degree of atomic order and no visible surface damage. The comparable $Q_\text{TLS}$ between these samples, despite their different surface quality, aligns with our earlier scaling analysis showing these BAW devices operate in a regime where bulk rather than surface TLS effects dominate.

Closer examination of the "Ion-mill + Anneal" sample reveals elongation of Nb atomic columns near the surface, suggesting localized lattice distortions. One possible explanation for these distortions is the presence of antisite defects, where Nb atoms substitute for Li sites. While the exact formation mechanism remains uncertain, it is plausible that a Li-deficient environment during annealing promotes Nb substitution at Li sites.
The substitution of Nb5+ for Li+ introduces both size and charge mismatch ($r_{Nb}=64~\text{pm}$ $r_{Li}=76~\text{pm}$). This mismatch likely induces significant local lattice strain and electrostatic imbalance, forcing the surrounding atoms to adjust. As a result, Nb atoms may not sit exactly in the Li sites but instead shift slightly offsite. Such deviations could manifest as column elongation and distortions observed in the HAADF-STEM images. To investigate these distortions further, we analyzed the ellipticity of atomic columns, where ellipticity arises when atoms shift within the plane perpendicular to the electron beam, causing otherwise circular atomic columns to appear elliptical in projection. The ellipticity analysis, presented in Appendix Figure \ref{fig:si_tem_ellipticity}, quantifies the magnitude and directionality of these distortions. We note that previous TLS studies in LN revealed no improvement of TLS loss after annealing in SAWs \cite{gruenke2024surface}; we propose the Nb elongation as a possible explanation for this lack of TLS loss improvement, even with removal of amorphous material via annealing.  
While these findings highlight the potential role of antisite defects in inducing localized lattice distortions, further experiments are necessary to confirm their presence and understand their origin. Ptychography, with its depth-resolved imaging capabilities and ability to simultaneously image light (e.g., Li and O) and heavy (e.g., Nb) elements, offers a promising pathway to directly image Li and O vacancies, NbLi antisites, and associated lattice distortions. This approach will allow for a more comprehensive understanding of the defect landscape.

\subsection{Atomic Force Microscopy}

To look at roughness and surface uniformity, we measure witness devices with equivalent surface treatments using atomic force microscopy (AFM). Each device is scanned several times with a Bruker Icon AFM in PeakForce mode to obtain 500nm square topography images. Roughness is extracted from each scan after a linear polynomial leveling of the image. The results are shown in Figure~\ref{fig:fig_afm}. Morphology is consistent across several scans of the same surface treated devices. The roughness across multiple scans of the same devices is similarly quite consistent; we calculate error for roughness from the standard deviation of all scans, which is shown as error bars in Figure~\ref{fig:fig_afm} (g). 

Morphology and roughness demonstrate significant surface changes through our surface treatment steps. After RIE cleaning, the surface is covered with debris, leaving a surface roughness that is much larger than standard chemical-mechanical polished (CMP) LN. After SC1 clean, this surface lying debris is removed, and the surface roughness is dramatically reduced. The surface topography of SC1 cleaned LN is similar to standard CMP polished LN, with amorphous a-periodic pitting on the surface. The "Ion mill" device (which includes an additional argon ion milling and SC1 clean from the "SC1" device) maintains the surface morphology while slightly reducing the surface roughness. 

Annealing at 500$^{\circ}$C for 72 hours followed by a piranha dip changes the surface morphology and reduces roughness, with or without ion mill+SC1 as an additional step. By annealing the material, the surface relaxes into atomically flat layers or terraces \cite{sanna2010lithium, sanna2014unraveling}. The terraces are observed to have larger spatial uniformity than previous LN annealing results for only 8 hours at the same temperature \cite{gruenke2024surface}. Allowing the material to anneal for longer time allows for better surface relaxation. Similarly, we notice that when ion milling+SC1 was included before annealing, the terraces are smaller and less uniform than when no ion milling was added. When compared with the TEM results showing an increased depth in amorphous damage layer prior to annealing, we suspect the extra ion mill damage requires longer annealing time to create the equivalent terrace uniformity. We also do not see a change in morphology after cooldown. We measure the same "Ion mill+Anneal" device before and after thermal cycling to 10~mK, which is seen in Figure~\ref{fig:fig_afm} (e) and (f). The terraces are maintained and have a similar size and uniformity. All samples that had annealing also had the lowest surface roughness.

\section{Results and Future Experiments}

In this study, we have demonstrated ultra-low loss BAW cavity resonators using bulk X-cut LN, with reproducible high power $Q_i$ above $10^7$. These devices have given us a unique window into TLS physics in LN. Despite major changes in surface quality from various treatments, we observe that TLS losses remain remarkably constant. This surprising result motivated us to compare BAW TLS loss measurements with previous work with LN SAWs and PNCs, allowing us to cleanly separate bulk and surface TLS contributions in LN for the first time. 

Our analysis reveals a clear regime where surface-lying amorphous LN and surface-lying defects and contamination dominate device performance - specifically for devices with surface calculated filling fraction above $10^{-5}$ per nanometer of damage layer. We establish quantitative bounds on these surface effects, finding surface loss tangent times damage layer depth falls between $1.05\times10^{-2} \pm 4.7\times10^{-3}$. This provides clear guidance for future device engineering: either reduce the surface participation ratio below $10^{-5}$ per nanometer, or improve surface quality. Future work with TEM using EELS and EDS will be valuable to understand the stoichiometry of the surface after etching and annealing, potentially revealing new ways to reduce surface loss. 

Our BAW measurements reveal a fundamental bulk TLS limit in LN, with $F\delta^0_{\text{TLS}}$ ranging from $8.78\times10^{-8} \pm 6.7\times10^{-9}$. We hypothesize these losses originate from lithium deficiency introduced during Czochralski growth, which creates niobium antisite defects throughout the material. Future investigations of BAW cavities fabricated from LN grown by different methods, including vapor transport equilibration, stoichiometric growth techniques, and precise dopant control, will be essential for understanding and minimizing bulk TLS losses in these promising quantum acoustic devices.

\begin{acknowledgments}
The authors would like to thank Dr. Christina Newcomb,  Kevin K. S. Multani, Oliver Hitchcock, Matthew Maksymowych, and Professor Martin Fejer experimental support and helpful discussions. We acknowledge support from the U.S. government through: the National Science Foundation CAREER award No. ECCS-1941826; the Office of Naval Research (ONR) under grant No. N00014-20-1-2422; the Air Force Office of Scientific Research and Office of Naval Research under award number FA9550-23-1-0338; the U.S. Air Force Office of Scientific Research (MURI Grant No. FA9550-17-1-0002); the U.S. Army Research Office (ARO)/Laboratory for Physical Sciences (LPS) Modular Quantum Gates (ModQ) program (Grant No. W911NF-23-1-0254); and the U.S. Department of Energy through Grant No. DE-AC02-76SF00515 (through SLAC) and via the Q-NEXT Center. We also acknowledge funding from Amazon Web Services Inc. and the Moore Inventor Fellowship. E.A.W. was supported by the Department of Defense through the National Defense \& Engineering Graduate Fellowship. Part of this work was performed at the Stanford Nano Shared Facilities (SNSF) and at the Stanford Nanofabrication Facility (SNF), supported by the National Science Foundation under award ECCS-2026822. The authors wish to thank NTT Research for their financial and technical support. Electron microscopy was performed at the Center for Electron Microscopy and Analysis (CEMAS) at The Ohio State University. S.S.-R. acknowledges Dan Huber for assisting with TEM sample preparation.
\end{acknowledgments}

\appendix

\section{Flat-Flat BAW design} \label{flatflat}
The BAW resonator is designed as a flip-chip packaged circuit: the bottom substrate containing the feedline and antenna necessary for driving the BAW waves, and the top substrate coupled non-galvanically to the antenna. This allows us to keep the substrate that hosts our measured BAW modes free from lift-off contamination. The feedline and antenna drive BAW modes in the bottom substrate as well. To assure we measure the longitudinally polarized BAW modes in the top surface treated substrate, we look for an FSR that matches the cavity length of the top substrate. 

\begin{figure}[ht]
\includegraphics[width=\linewidth]{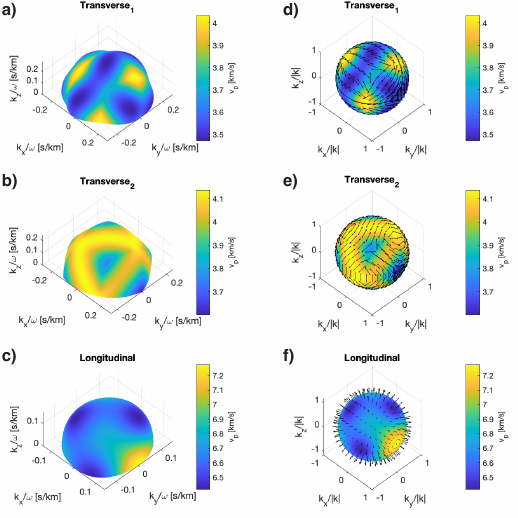}
\caption{\label{fig:si_slowness} \emph{Slowness Surface} Using the Christoffel equation, we plot the slowness surface of X-cut LN for the three polarizations of bulk waves. a-c) The slowness surface for arbitrary direction of propagation in LN. Note, these slowness surfaces have been rotated to reflect X-cut LN, such that propagation along crystal $\hat{x}$ corresponds to $(k_x,k_y) = 0$. The color corresponds to the phase velocity of the acoustic wave. d-f) We plot the surface of the normalized direction of propagation as a function of phase velocity (color) and polarization direction (black vectors). }
\end{figure}

\begin{figure}[ht]
\includegraphics[width=\linewidth]{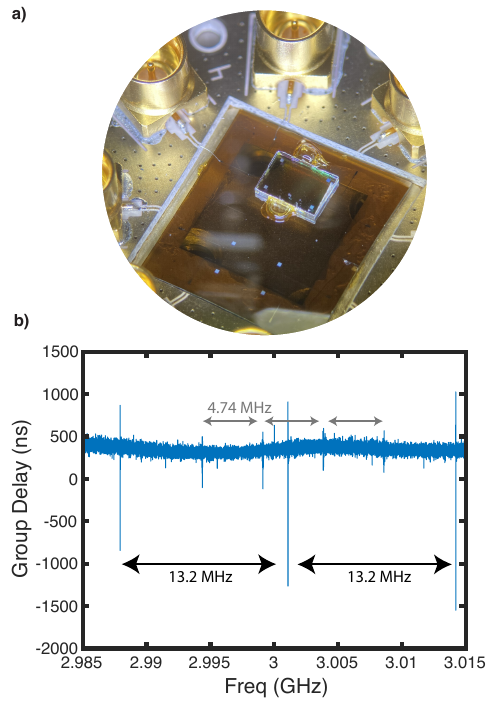}
\caption{\label{fig:si_groupdelay} \emph{Device Image and Measured Group Delay} a) We show an image of the device, packaged with flip-chip architecture. The bottom substrate has patterned antenna electrodes, tied to SMP connectors on our PCB via a feedline and wirebonds. The top substrate hosts the longitudinal BAW modes. We place antennas near the edge of the top substrate, such that the feedline does not drive other BAW polarizations in the top substrate. The two chips are secured with a GE varnish adhesive. b) A plot of the group delay measured on reflection of our device. Two unique FSRs are identified. The 13.2 MHz FSR corresponds to the longitudinal BAW modes in the top chip cavity; the other corresponds to transverse BAW modes driven in the bottom chip cavity. By identifying the correct FSR, we isolate which modes to probe for the TLS surface-treatment study. }
\end{figure}

We solve the Christoffel equation to find the slowness surfaces for x-cut LN \cite{vonLupke2023thesis, jaeken2016solving} (see Figure~\ref{fig:si_slowness}). The longitudinal speed of sound for LN traveling along crystal $\hat{x}$ is 6622 km/s, while the speed of sound for the two transverse polarizations is 3472 and 4137 m/s. For a cavity length of $L = 500 \mu m$, we expect a longitudinal FSR of $v/2L = 6.6$ MHz. A plot of the group delay is shown in Figure~\ref{fig:si_groupdelay} (b). We observe an FSR of 13.2 MHz, agreeing well with twice the longitudinal FSR. We suspect that this is due to weak coupling between the fairly uniform electric field throughout the bulk and the open strain boundary conditions on both sides of the cavity: the coupling to an even $m$-number longitudinal mode's strain field cancels in a uniform electric field. We also observe an FSR of 4.74 MHz, corresponding to transverse modes propagating in the bottom metalized substrate. Unlike the longitudinal modes where the uniform electric field throughout the bulk selectively drives odd $m$-number modes, the transverse modes show no such selectivity likely due to the different boundary conditions at the surfaces of the eletrode chip which reduce the effect of the cancellation. 

\section{Acoustic Beam Diffraction Loss}
\label{diffraction}

The high power $Q_i$ of our BAW resonators from $0.1-10$~GHz  increases up to frequencies around 8~GHz before it flattens with a $Q_i$ above $10^7$. This behavior immediately suggests that losses due to Rayleigh scattering in bulk or on the surface which increase at higher frequencies are not the dominating loss channel in our BAW resonators. Both Rayleigh losses and mirror roughness losses increase with frequency, such that the $Q_{i,res}$ should decrease frequency \cite{goryachev2013observation, straniero2015realistic}, and we consider this in more detail in Appendix~\ref{app:rayleigh}.

The losses that increase at lower frequencies are likely caused by diffraction of the acoustic beam inside the resonator. We simulate this by using a Fox and Li style beam propagation simulation 
\cite{newberry1989paraxial, renninger2018bulk, vonLupke2023thesis, fox1961resonant}. By initializing two infinite radius of curvature acoustic mirrors, the slowness surface for X-cut LN, the length of our cavity ($L = 500$ microns), the frequency of beam propagation, and a Gaussian input acoustic displacement field $u_0$, we solve for the acoustic displacement field at the mirror surface after one round trip through the cavity $u_1$. From the acoustic displacement field we solve for the displacement overlap $\eta$ after one round trip through the cavity by, 
\begin{equation}
    \eta = \frac{\int u_0^* u_1}{\sqrt{\int|u_0|^2 \int|u_1|^2}}.
\end{equation} From the mode overlap, we calculate the round trip loss $1-\eta^2$ and the resulting quality factor $Q = 2 \pi f \tau_{RT}/ (1-\eta^2)$, where $\tau_{RT}$ is the round trip travel time of the acoustic beam. We sweep this calculation over propagation frequency, and therefore find the diffraction-limited internal quality factor as a function of frequency. We also bound the upper limit of the quality factor above 10 GHz by the high power TLS losses. This model for losses matches well with our measured device $Q_i$, seen in Figure~\ref{fig:fig_losses}. 

\section{Rayleigh Scattering Loss Estimate\label{app:rayleigh}}

As mentioned above, lack of a clear $\omega^{-3}$ scaling of the quality factor suggests that Rayleigh scattering does not limit the losses of our BAW resonators. To estimate the limits imposed by Rayleigh scattering in our lithium niobate BAW resonators, we begin with Klemens' formula for scattering from point defects \cite{klemens1955scattering}: $\tau^{-1} = \frac{a^3}{G} (\frac{\Delta M}{M})^2 \frac{\omega^4}{4\pi v^3}$, where $a$ is the lattice constant, $G$ is the total number of atoms in the crystal (such that $G a^3$ is the crystal volume), $M$ is the mass of each atom, and $\Delta M$ is the mass difference introduced by the defect. For a system with multiple independent scatterers with density $N$, the total scattering rate is $\tau_{\text{tot}}^{-1} = N V \tau^{-1}$, leading to a quality factor limit of $Q^{-1}_{\text{Rayleigh}} = \frac{N \omega^3}{4\pi v^3} \cdot \frac{m^2}{\rho^2}$, where $m$ is the defect mass difference and $\rho$ is the material density.

In congruent lithium niobate, approximately 3.1\% of lithium atoms are replaced by niobium antisite defects. Using the atomic masses of Li (6.941 g/mol) and Nb (92.906 g/mol), this gives a mass difference per defect of $m \approx 1.43 \times 10^{-25}$ kg. Given the density of LiNbO$_3$ ($\rho = 4640~\mathrm{kg/m}^3$) and its molecular mass (147.85 g/mol), we calculate a defect density of $N \approx 5.86 \times 10^{26}~\mathrm{m}^{-3}$ --- roughly a defect every cubic nanometer. With these parameters and assuming a longitudinal sound velocity of roughly 6600 m/s, we estimate $Q_{\text{Rayleigh}} \approx 2.7 \times 10^{10}$ at 10 GHz. Though this is a rough estimate, and may be off by an order of magnitude, it is over two orders of magnitude larger than our measured quality factors at 10 GHz, and four orders of magnitude larger than quality factors at 3 GHz, indicating that Rayleigh scattering from mass-mismatch defects is not the dominant loss mechanism in our devices, and will only limit devices at frequencies exceeding $10$ GHz. Finally we note that our estimates (and our experimental results) suggest that Rayleigh scattering due to part per billion level atomic impurities do not lead to the losses observed in Ref.~\cite{goryachev2013observation}.

\section{Calculating Surface Filling Fraction}
\label{calc_F}

To compare the LN PNCs, SAWs, and BAWs, we calculate the filling fraction due to a thin surface layer of thickness $d$, or $F_\text{surface}$. The filling fraction is defined by Equation \ref{eq:fill_gen}. For the resonators that are simulated with FEM simulations (the PNCs and the SAWs), we can calculate $F_\text{surface}$ from COMSOL, by defining a domain of depth $d$ and computing the acoustic energy in the surface domain normalized by the entire resonator acoustic energy. 

However, our BAW devices are not simulated by COMSOL due to computational constraints. Therefore, we calculate a surface filling fraction analytically from the strain. We assume the BAW has a strain profile of $S_{xx} = S_0 LG_{p l}(r,\phi) \cos{\frac{m \pi z}{L}}$, where $S_0$ is a scaling factor and $LG_{p l}$ are the Laguerre-Gaussian modes, and $L$ is the length of the BAW cavity \cite{vonLupke2023thesis}. We assume the strain profile is of $\cos{\frac{m \pi z}{L}}$ because both mirrors have an open boundary condition. With a form for the dominant strain component in the BAW, we calculate the ratio of the acoustic energy in a volume of TLS material normalized by the total acoustic energy in all the BAW material. Because we define the TLS hosting material as a uniform surface layer with small thickness $d$, the TLS hosting volume is constant across the top plane of the BAW wave. Therefore, $F_\text{surface}$ is independent of the transverse strain profile. The calculation of $F_\text{surface}$ for our longitudinal BAW modes is shown below: 

\begin{equation}
    \begin{split}
        F &= \frac{\int_{r,\phi} \int_{z_h} |S_{xx}|^2 dV_h}{\int_{r,\phi} \int_{z} |S_{xx}|^2 dV} \\
         &= \frac{\int_0^{d} |\cos{\frac{m \pi z}{L}|^2 dz}}{\int_0^L |\cos{\frac{m \pi z}{L}|^2 dz}} \\
         &= (\frac{d}{L} + \frac{ \sin{\frac{2 m \pi d}{L}}}{2 m \pi}) \approx \frac{2d}{L} \\
    \end{split}
\end{equation}

\section{TLS Loss Independence} \label{TLS_independence}

When comparing resonators of different surface treatments, we want to ensure that the measured $F\delta^0_{\text{TLS}}$ is independent of all other changes that occur between devices. We note that because these devices have no lensing from a curved mirror surface, the mode family supports many modes that often change coupling from device to device. This means the highest SNR modes may not always have the same planar acoustic mode shape when chosen between different devices. Because $F$ is normalized by the acoustic energy by volume and the TLS volume is assumed to be uniformly distributed across the surface, changes in the transverse mode profile do not affect $F\delta^0_{\text{TLS}}$. On two control "RIE" devices, we temperature sweep all acoustic modes in the mode family at 4209 MHz. From our extracted $F\delta^0_{\text{TLS}}$, we see experimentally that there is very little variation in TLS loss across the mode family. The maximum measured change in $F\delta^0_{\text{TLS}}$ from different modes of the mode family of $8\times10^{-9}$ , less than both the total error measured from a single device between 4 and 8 GHz as well as the differences between surface treated devices. We therefore conclude that changes from choosing different modes in the mode family does not impact results of the surface treatment study. 

\begin{figure}[ht]
\includegraphics[width=\linewidth]{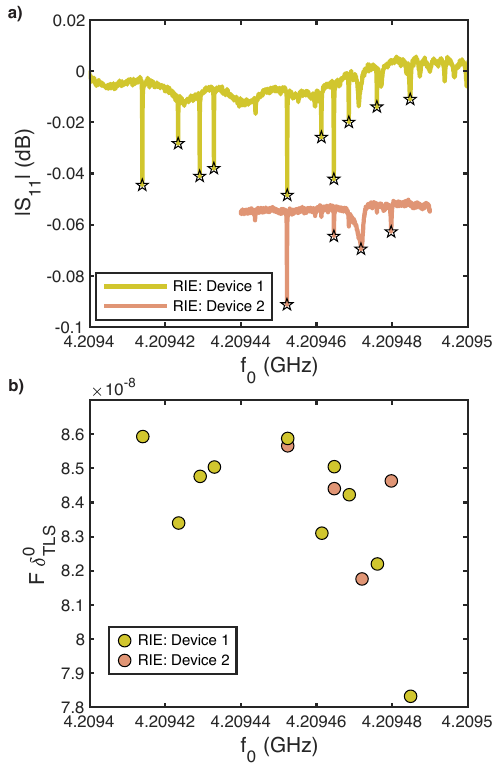}
\caption{\label{fig:si_f_modefamily} \emph{$F\delta^0_{\text{TLS}}$ vs Mode Family} For two control devices, we measure all resonances in the mode family centered around 4209 MHz for their temperature dependent TLS losses. a) The spectrum of the full mode family at 4209 MHz for two "RIE" control devices. Each resonance that is measured for TLS losses is starred. b) The extracted TLS losses for each resonance in the mode family. We observe little variation in the TLS loss from mode to mode, showing a lack of radial strain dependence on the TLS losses. This confirms that we can compare the TLS losses between surface treated resonators even when the mode families differ between devices.}
\end{figure}

Similarly, changes in coupling to single modes in the mode family as well as differences in $Q_{i,res}$ change the SNR and internal phonon number for a given input power. While we measure all modes of all surface treated devices with a average internal phonon number between $10^8 < \langle n \rangle < 10^9$ (or similarly an estimated $10^2 < \langle \frac{\langle n \rangle}{n_c} \rangle < 10^3$), we also evaluate for a single resonator the impact of the extracting TLS loss for various input powers. On one "Ion Mill" device, we measure the temperature dependent TLS loss for four resonances at three different powers. While the changes in extract $F\delta^0_{\text{TLS}}$ is greater than the changes caused by various mode families or seen from device to device, we note there are no consistent trends in the power dependent $F\delta^0_{\text{TLS}}$ across the four resonance frequencies. We therefore assume that the changes induced by different powers are caused by increased fit error for lower SNR (lower power) measurements, rather than by true TLS loss changes as a function of power. We again conclude that our TLS measurements are not impacted by a factor of 10 difference in $\frac{\langle n \rangle}{n_c}$, so long as the SNR is high.  

\begin{figure}[ht]
\includegraphics[width=\linewidth]{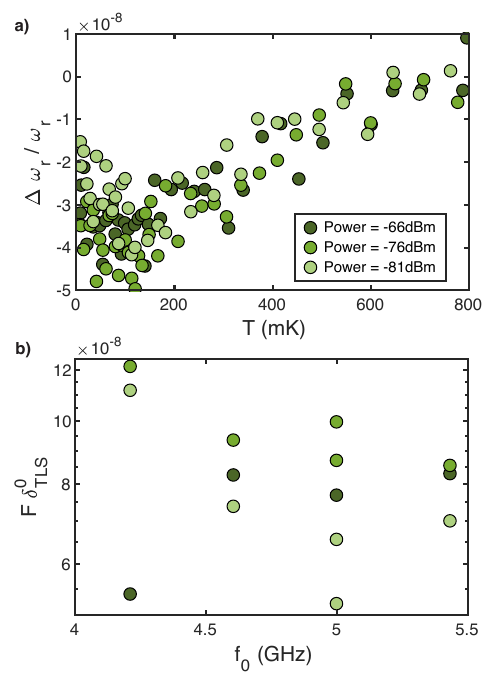}
\caption{\label{fig:si_f_pow} \emph{$F\delta^0_{\text{TLS}}$ vs Power} We repeat temperature swept TLS loss measurements for our "Ion mill" device on multiple resonances between 4 and 5.5 GHz and at multiple powers. a) We show the temperature dependent resonant frequency shift for the resonance at 4603MHz at three unique powers. b) The extracted $F\delta^0_{\text{TLS}}$ for each resonance at each measured power. We show no trend of decreasing or increasing $F\delta^0_{\text{TLS}}$ with power.}
\end{figure}

\section{Ellipticity Analysis of Atomic Columns} \label{TEM_ellipticity}

\begin{figure*}[ht]
\includegraphics[width=\linewidth]{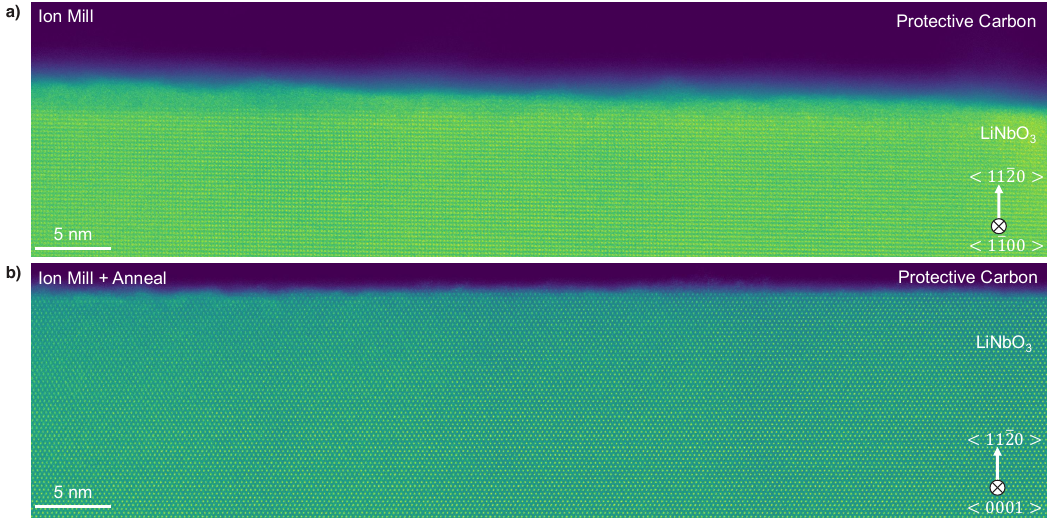}
\caption{\label{fig:si_tem_highres} \emph{High resolution TEMs} HAADF-STEM images of cross-sectioned "Ion mill" and "Ion-mill + Anneal" lithium niobate samples. The "Ion mill" sample (top) shows a damaged, amorphous surface layer approximately 3 nm thick, while the "Ion mill + Anneal" sample (bottom) exhibits a restored, well-ordered crystalline structure. A protective Sharpie carbon layer was applied prior to the FIB process to prevent surface damage during sample preparation.}
\end{figure*}

\begin{figure*}[ht]
\includegraphics[width=\linewidth]{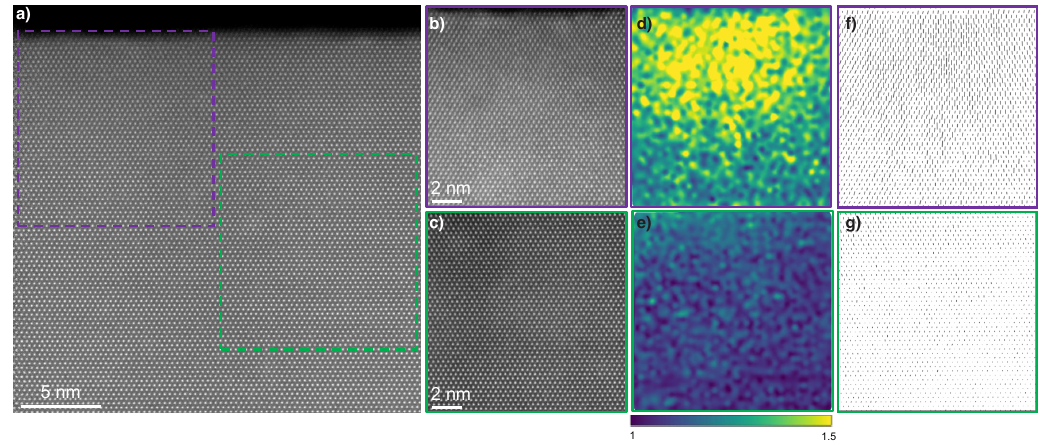}
\caption{\label{fig:si_tem_ellipticity} \emph{TEM Ellipticity} HAADF-STEM images and ellipticity analysis of atomic columns near the surface of the "Ion mill + Anneal" sample. (a) HAADF-STEM image showing regions of interest analyzed for ellipticity. (b, c) Zoomed-in HAADF-STEM images of the selected regions, marked by purple and green boxes, respectively. (d,e) Ellipticity maps showing the magnitude of column distortions, with brighter regions corresponding to higher ellipticity. (f, g) Vector plots indicating the magnitude and directionality of ellipticity across the analyzed regions. The observed column distortions near the surface are consistent with localized lattice strain, possibly induced by antisite defects.}
\end{figure*}

To investigate the nature of lattice distortions near the surface of the "Ion mill + Anneal" sample, we performed ellipticity analysis of the atomic columns. Ellipticity measures deviations from circularity in atomic columns and arises when atoms shift within the plane perpendicular to the
electron beam, causing circular columns to appear elliptical in projection. Perfectly circular columns have an ellipticity value of 1, while values greater than 1 indicate increasing elliptical distortions. The results of the ellipticity analysis are shown in Supplementary Figure~\ref{fig:si_tem_ellipticity}. The HAADF-STEM image (a–c) highlights the regions of interest for analysis. The ellipticity maps (d, e) display the magnitude of distortions, with brighter regions corresponding to higher ellipticity values. These distortions are most pronounced closer to the surface, consistent with localized lattice strain. The vector plots (f, g) provide additional insights, illustrating both the magnitude and directionality of ellipticity through arrow length and orientation. These observations suggest that Nb atomic columns near the surface experience measurable lattice distortions. One plausible explanation is the presence of antisite defects. This substitution introduces both size and charge mismatch, creating local strain fields that force atomic positions to shift slightly offsite. These shifts   manifest   as   the   observed   column   elongation   and   ellipticity   in   projection. Further experimental validation is needed to confirm the exact nature and origin of these defects. Ptychography combined with simulation-assisted imaging, will help clarify the role of antisite defects in lattice distortions and their potential contribution to TLS losses.

\clearpage

\bibliography{bib}

\end{document}